\documentclass{svmult}

\usepackage{makeidx}         
\usepackage{graphicx}        
\usepackage{multicol}        
\usepackage{color}
\usepackage{color}
\usepackage{multirow}

\makeindex             

\begin{document}

\title*{The Tyranny of Data?\\The Bright and Dark Sides of Data-Driven Decision-Making for Social Good}
\author{Bruno Lepri, Jacopo Staiano, David Sangokoya, Emmanuel Letouz\'{e} and Nuria Oliver}
\institute{Bruno Lepri \at Fondazione Bruno Kessler 
\email{lepri@fbk.eu}
\and Jacopo Staiano \at Fortia Financial Solutions
\email{jacopo.staiano@fortia.fr}
\and David Sangokoya \at Data-Pop Alliance \email{dsangokoya@datapopalliance.org}
\and Emmanuel Letouz\'{e} \at Data-Pop Alliance and MIT Media Lab
\email{eletouze@mit.edu}
\and Nuria Oliver \at Data-Pop Alliance \email{nuria@alum.mit.edu}}
%
%
\maketitle

\abstract{The unprecedented availability of large-scale human behavioral data is profoundly changing the world we live in. Researchers, companies, governments, financial institutions, non-governmental organizations and also citizen groups are actively experimenting, innovating and adapting algorithmic decision-making tools to understand global patterns of human behavior and provide decision support to tackle problems of societal importance. In this chapter, we focus our attention on \emph{social good decision-making algorithms}, that is algorithms strongly influencing decision-making and resource optimization of public goods, such as public health, safety, access to finance and fair employment. Through an analysis of specific use cases and approaches, we highlight both the positive opportunities that are created through data-driven algorithmic decision-making, and the potential negative consequences that practitioners should be aware of and address in order to truly realize the potential of this emergent field. We elaborate on the need for these algorithms to provide transparency and accountability, preserve privacy and be tested and evaluated \emph{in context}, by means of living lab approaches involving citizens. Finally, we turn to the requirements which would make it possible to leverage the predictive power of data-driven human behavior analysis while ensuring transparency, accountability, and civic participation.}

\section{Introduction}
\label{sec:1}
The world is experiencing an unprecedented transition where human behavioral data has evolved from being a scarce resource to being a massive and real-time stream. This availability of large-scale data is profoundly changing the world we live in and has led to the emergence of a new discipline called \emph{computational social science} \cite{Lazer09}; finance, economics, marketing, public health, medicine, biology, politics, urban science and journalism, to name a few, have all been disrupted to some degree by this trend \cite{king2011ensuring}.

Moreover, the automated analysis of anonymized and aggregated large-scale human behavioral data offers new possibilities to understand global patterns of human behavior and to help decision makers tackle problems of societal importance \cite{Lazer09}, such as monitoring socio-economic deprivation \cite{Blumenstock2015,Capra2014,Soto2011,venerandi2015} and crime \cite{bogomolov2014,bogomolov2015,Toole2011,traunmueller2014,wang2016}, mapping the propagation of diseases \cite{ginsberg2009, Wesolowski2012}, or understanding the impact of natural disasters \cite{ofli2016,pastor2014,Wilson16}. Thus, researchers, companies, governments, financial institutions, non-governmental organizations and also citizen groups are actively experimenting, innovating and adapting algorithmic decision-making tools, often relying on the analysis of personal information.

However, researchers from different disciplinary backgrounds have identified a range of social, ethical and legal issues surrounding data-driven decision-making, including privacy and security \cite{crawfordSchultz2014,deMontjoye2013,deMontjoye2015,ohm2010}, transparency and accountability \cite{citronPasquale2014,pasquale2015,zarsky2016,zarsky2012}, and bias and discrimination \cite{barocas2016,sweeney13}. For example, Barocas and Selbst \cite{barocas2016} point out that the use of data-driven decision making processes can result in disproportionate adverse outcomes for disadvantaged groups, in ways that look like discrimination. Algorithmic decisions can reproduce patterns of discrimination, due to decision makers' prejudices \cite{pager2008}, or reflect the biases present in the society \cite{pager2008}. In 2014, the White House released a report, titled ``Big Data: Seizing opportunities, preserving values'' \cite{whitehouse} that highlights the discriminatory potential of big data, including how it could undermine longstanding civil rights protections governing the use of personal information for credit, health, safety, employment, etc.
For example, data-driven decisions about applicants for jobs, schools or credit may be affected by hidden biases that tend to flag individuals from particular demographic groups as unfavorable for such opportunities. Such outcomes can be self-reinforcing, since systematically reducing individuals' access to credit, employment and educational opportunities may worsen their situation, which can play against them in future applications.

In this chapter, we focus our attention on \emph{social good algorithms}, that is algorithms strongly influencing decision-making and resource optimization of public goods, such as public health, safety, access to finance and fair employment. These algorithms are of particular interest given the magnitude of their impact on quality of life and the risks associated with the information asymmetry surrounding their governance.

In a recent book, William Easterly evaluates how global economic development and poverty alleviation projects have been governed by a ``tyranny of experts'' -- in this case, aid agencies, economists, think tanks and other analysts -- who consistently favor top-down, technocratic governance approaches at the expense of the individual rights of citizens \cite{Easterly2014}. Easterly details how these experts reduce multidimensional social phenomena such as poverty or justice into a set of technical solutions that do not take into account either the political systems in which they operate or the rights of intended beneficiaries. Take for example the displacement of farmers in the Mubende district of Uganda: as a direct result of a World Bank project intended to raise the region's income by converting land to higher value uses, farmers in this district were forcibly removed from their homes by government soldiers in order to prepare for a British company to plant trees in the area \cite{Easterly2014}. Easterly underlines the cyclic nature of this tyranny: technocratic justifications for specific interventions are considered objective; intended beneficiaries are unaware of the opaque, black box decision-making involved in these resource optimization interventions; and experts (and the coercive powers which employ them) act with impunity and without redress. 

If we turn to the use, governance and deployment of big data approaches in the public sector, we can draw several parallels towards what we refer to as the ``tyranny of data", that is the adoption of data-driven decision-making under the technocratic and top-down approaches higlighted by Easterly \cite{Easterly2014}. We elaborate on the need for \emph{social good decision-making algorithms} to provide transparency and accountability, to only use personal information -- owned and controlled by individuals -- with explicit consent, to ensure that privacy is preserved when data is analyzed in aggregated and anonymized form, and to be tested and evaluated \emph{in context}, that is by means of living lab approaches involving citizens. In our view, these characteristics are crucial for fair data-driven decision-making as well as for citizen engagement and participation.

In the rest of this chapter, we provide the readers with a compendium of the issues arising from current big data approaches, with a particular focus on specific use cases that have been carried out to date, including urban crime prediction \cite{bogomolov2015}, inferring socioeconomic status of countries and individuals \cite{Blumenstock2015,eagle2014,Soto2011}, mapping the propagation of diseases \cite{ginsberg2009, Wesolowski2012} and modeling individuals' mental health \cite{bogomolov2014stress,deChoudhury2013,likamwa2013}. Furthermore, we highlight factors of risk (\emph{e.g.} privacy violations, lack of transparency and discrimination) that might arise when decisions potentially impacting the daily lives of people are heavily rooted in the outcomes of black-box data-driven predictive models. Finally, we turn to the requirements which would make it possible to leverage the predictive power of data-driven human behavior analysis while ensuring transparency, accountability, and civic participation. 

\section{The rise of data-driven decision-making for social good}

The unprecedented stream of large-scale, human behavioral data has been described as a ``tidal wave'' of opportunities to both predict and act upon the analysis of the petabytes of digital signals and traces of human actions and interactions. With such massive streams of relevant data to mine and train algorithms with, as well as increased analytical and technical capacities, it is of no surprise that companies and public sector actors are turning to machine learning-based algorithms to tackle complex problems at the limits of human decision-making \cite{Gillespie2014, Willson2016}. The history of human decision-making -- particularly when it comes to questions of power in resource allocation, fairness, justice, and other public goods -- is wrought with innumerable examples of extreme bias, leading towards corrupt, inefficient or unjust processes and outcomes \cite{akerlof2009,fiske1998,samuelson1988,tversky1974}. In short, human decision-making has shown significant limitations and the turn towards data-driven algorithms reflects a search for objectivity, evidence-based decision-making, and a better understanding of our resources and behaviors.

Diakopoulos \cite{Diakopoulos2015} characterizes the function and power of algorithms in four broad categories: 1)  \emph{classification}, the categorization of information into separate ``classes'', based on its features; 2) \emph{prioritization}, the denotation of emphasis and rank on particular information or results at the expense of others based on a pre-defined set of criteria; 3) \emph{association}, the determination of correlated relationships between entities; and 4) \emph{filtering}, the inclusion or exclusion of information based on pre-determined criteria. 

Table~\ref{tab:algorithms} provides examples of types of algorithms across these categories.

\begin{table}[]
\centering
\caption{Algorithmic function and examples, adapted from Diakopoulos \cite{Diakopoulos2015} and Latzer \emph{et al.} \cite{Latzer2015}}
\label{tab:algorithms}
\begin{tabular}{lll}
\hline
\multicolumn{1}{|l|}{\textbf{Function}} & \multicolumn{1}{l|}{\textbf{Type}} & \multicolumn{1}{l|}{\textbf{Examples}} \\ \hline
\textit{Prioritization} & \begin{tabular}[c]{@{}l@{}}General and search engines, \\ meta search engines, semantic \\ search engines, questions \& \\ answers services\end{tabular} & \begin{tabular}[c]{@{}l@{}}Google, Bing, Baidu;\\ image search; social\\ media; Quora; Ask.com\end{tabular} \\ \hline
\textit{Classification} & \begin{tabular}[c]{@{}l@{}}Reputation systems, news scoring, \\ credit scoring, social scoring\end{tabular} & \begin{tabular}[c]{@{}l@{}}Ebay, Uber, Airbnb;\\ Reddit, Digg; \\ CreditKarma; Klout\end{tabular} \\ \hline
\textit{Association} & \begin{tabular}[c]{@{}l@{}}Predicting developments and\\ trends\end{tabular} & \begin{tabular}[c]{@{}l@{}}ScoreAhit, Music Xray, \\ Google Flu Trends\end{tabular} \\ \hline
\textit{Filtering} & \begin{tabular}[c]{@{}l@{}}Spam filters, child protection filters,\\ recommender systems, news\\ aggregators\end{tabular} & \begin{tabular}[c]{@{}l@{}}Norton; Net Nanny;\\ Spotify, Netflix;\\ Facebook Newsfeed\end{tabular}
\end{tabular}
\end{table}

This chapter places emphasis on what we call social good algorithms -- algorithms strongly influencing decision-making and resource optimization for public goods. These algorithms are designed to analyze massive amounts of human behavioral data from various sources and then, based on pre-determined criteria, select the information most relevant to their intended purpose. While resource allocation and decision optimization over limited resources remain common features of the public sector, the use of social good algorithms brings to a new level the amount of human behavioral data that public sector actors can access, the capacities with which they can analyze this information and deliver results, and the communities of experts and common people who hold these results to be objective. The ability of these algorithms to identify, select and determine information of relevance beyond the scope of human decision-making creates a new kind of decision optimization faciliated by both the design of the algorithms and the data from which they are based. However, as discussed later in the chapter, this new process is often opaque and assumes a level of impartiality that is not always accurate. It also creates information asymmetry and lack of transparency between actors using these algorithms and the intended beneficiaries whose data is being used. 

In the following sub-sections, we assess the nature, function and impact of the use of social good algorithms in three key areas: criminal behavior dynamics and predictive policing; socio-economic deprivation and financial inclusion; and public health. 

\subsection{Criminal behavior dynamics and predictive policing}
Researchers have turned their attention to the automatic analysis of criminal behavior dynamics both from a \emph{people}- and a \emph{place}-centric perspectives. The people-centric perspective has mostly been used for individual or collective criminal profiling \cite{ratcliffe2006,Short2008,wang2013}. For example, Wang \emph{et al.} \cite{wang2013} proposed a machine learning approach, called Series Finder, to the problem of detecting specific patterns in crimes that are committed by the same offender or group of offenders.

In 2008, the criminologist David Weisburd proposed a shift from a people-centric paradigm of police practices to a place-centric one \cite{Weisburd2008}, thus focusing on geographical topology and micro-structures rather than on criminal profiling. An example of a place-centric perspective is the detection, analysis, and interpretation of crime hotspots \cite{Chainey2008,Eck2005,Mohler2011}. Along these lines, a novel application of quantitative tools from mathematics, physics and signal processing has been proposed by Toole \emph{et al.} \cite{Toole2011} to analyse spatial and temporal patterns in criminal offense records. Their analyses of crime data from 1991 to 1999 for the American city of Philadelphia indicated the existence of multi-scale complex relationships in space and time. Further, over the last few years, aggregated and anonymized mobile phone data has opened new possibilities to study city dynamics with unprecedented temporal and spatial granularities \cite{Blondel15}. Recent work has used this type of data to predict crime hotspots through machine-learning algorithms \cite{bogomolov2015,bogomolov2014,traunmueller2014}.

More recently, these \emph{predictive policing} approaches \cite{perry2013} are moving from the academic realm (universities and research centers) to police departments. In Chicago, police officers are paying particular attention to those individuals flagged, through risk analysis techniques, as most likely to be involved in future violence. In Santa Cruz, California, the police have reported a dramatic reduction in burglaries after adopting algorithms that predict where new burglaries are likely to occur. In Charlotte, North Carolina, the police department has generated a map of high-risk areas that are likely to be hit by crime. The Police Departments of Los Angeles, Atlanta and more than 50 other cities in the US are using PredPol, an algorithm that generates 500 by 500 square foot predictive boxes on maps, indicating areas where crime is most likely to occur. Similar approaches have also been implemented in Brasil, the UK and the Netherlands. Overall, four main predictive policing approaches are currently being used: (i) methods to forecast places and times with an increased risk of crime \cite{ferguson2012}, (ii) methods to detect offenders and flag individuals at risk of offending in the future \cite{perry2013}, (iii) methods to identify perpetrators \cite{perry2013}, and (iv) methods to identify groups or, in some cases, individuals who are likely to become the victims of crime \cite{perry2013}.

\subsection{Socio-economic deprivation and financial inclusion}
Being able to accurately measure and monitor key sociodemographic and economic indicators is critical to design and implement public policies \cite{ravallion2016}. For example, the geographic distribution of poverty and wealth is used by governments to make decisions about how to allocate scarce resources and provides a foundation for the study of the determinants of economic growth \cite{fields1989,kuznets1955}. The quantity and quality of economic data available have significantly improved in recent years. However, the scarcity of reliable key measures in developing countries represents a major challenge to researchers and policy-makers\footnote{\url{http://www.undatarevolution.org/report/}}, thus hampering efforts to target interventions effectively to areas of greatest need (\emph{e.g.} African countries) \cite{devarajan2013,jerven2013}.
Recently, several researchers have started to use mobile phone data \cite{Blumenstock2015,eagle2014,Soto2011}, social media \cite{venerandi2015} and satellite imagery \cite{jean2016} to infer the poverty and wealth of individual subscribers, as well as to create high-resolution maps of the geographic distribution of wealth and deprivation.

The use of novel sources of behavioral data and algorithmic decision-making processes is also playing a growing role in the area of financial services, for example credit scoring. Credit scoring is a widely used tool in the financial sector to compute the risks of lending to potential credit customers. Providing information about the ability of customers to pay back their debts or conversely to default, credit scores have become a key variable to build financial models of customers. Thus, as lenders have moved from traditional interview-based decisions to data-driven models to assess credit risk, consumer lending and credit scoring have become increasingly sophisticated. Automated credit scoring has become a standard input into the pricing of mortgages, auto loans, and unsecured credit. However, this approach is mainly based on the past financial history of customers (people or businesses) \cite{thomas2009}, and thus not adequate to provide credit access to people or businesses when no financial history is available. Therefore, researchers and companies are investigating novel sources of data to replace or to improve traditional credit scores, potentially opening credit access to individuals or businesses that traditionally have had poor or no access to mainstream financial services --\emph{e.g.} people who are unbanked or underbanked, new immigrants, graduating students, etc. 
Researchers have leveraged mobility patterns from credit card transactions \cite{Singh15} and mobility and communication patterns from mobile phones to automatically build user models of spending behavior \cite{Singh2013} and propensity to credit defaults \cite{sanpedro2015,Singh15}.
The use of mobile phone, social media, and browsing data for financial risk assessment has also attracted the attention of several entrepreneurial efforts, such as Cignifi\footnote{\url{http://cignifi.com/}}, Lenddo\footnote{\url{https://www.lenddo.com/}}, InVenture\footnote{\url{http://tala.co/}}, and  ZestFinance\footnote{\url{https://www.zestfinance.com/}}.

\subsection{Public health}
The characterization of individuals and entire populations' mobility is of paramount importance for public health \cite{oliver15}: for example, it is key to predict the spatial and temporal risk of diseases \cite{FriasMartinez2011,tizzoni2014,Wesolowski2012}, to quantify exposure to air pollution \cite{liu2013}, to understand human migrations after natural disasters or emergency situations \cite{bengtsson2011,lu2012}, etc. The traditional approach has been based on household surveys and information provided from census data. These methods suffer from recall bias and limitations in the size of the population sample, mainly due to excessive costs in the acquisition of the data. Moreover, survey or census data provide a snapshot of the population dynamics at a given moment in time. However, it is fundamental to monitor mobility patterns in a continuous manner, in particular during emergencies in order to support decision making or assess the impact of government measures.

Tizzoni \emph{et al.} \cite{tizzoni2014} and Wesolowski \emph{et al.} \cite{Wesolowski2014} have compared traditional mobility surveys with the information provided by mobile phone data (Call Detail Records or CDRs), specifically to model the spread of diseases. The findings of these works recommend the use of mobile phone data, by themselves or in combination with traditional sources, in particular in low-income economies where the availability of surveys is highly limited.

Another important area of opportunity within public health is mental health. Mental health problems are recognized to be a major public health issue\footnote{\url{http://www.who.int/topics/mental\_health/en/}}. However, the traditional model of episodic care is suboptimal to prevent mental health outcomes and improve chronic disease outcomes. In order to assess human behavior in the context of mental wellbeing, the standard clinical practice relies on periodic self-reports that suffer from subjectivity and memory biases, and are likely influenced by the current mood state. Moreover, individuals with mental conditions typically visit doctors when the crisis has already happened and thus report limited information about precursors useful to prevent the crisis onset. These novel sources of behavioral data yield the possibility of monitoring mental health-related behaviors and symptoms outside of clinical settings and without having to depend on self-reported information \cite{matic16}. For example, several studies have shown that behavioral data collected through mobile phones and social media can be exploited to recognize bipolar disorders \cite{deChoudhury2013,faurholt14,osmani2015}, mood \cite{likamwa2013}, personality \cite{deoliveira2011,lepri2016umuai} and stress \cite{bogomolov2014stress}.

Table~\ref{tab:summary} summarizes the main points emerging from the literture reviewed in this section.

\begin{table}[]
    \centering
    \caption{Summary table for the literature discussed in Section 2.}
    \label{tab:summary}
    \begin{tabular}{lll}
\hline
\multicolumn{1}{|l|}{\textbf{Key Area}} & \multicolumn{1}{l|}{\textbf{Problems Tackled}} & \multicolumn{1}{l|}{\textbf{References}} \\ \hline
\textit{Predictive Policing} & \begin{tabular}[c]{@{}l@{}}Criminal behavior profiling\\Crime hotspot prediction \\Perpetrator(s)/victim(s) identification\end{tabular} & \begin{tabular}[c]{@{}l@{}}\cite{ratcliffe2006,Short2008,wang2013}\\ \cite{bogomolov2015,bogomolov2014,ferguson2012,traunmueller2014}\\ \cite{perry2013}\end{tabular} \\ \hline
\textit{Finance \& Economy} & \begin{tabular}[c]{@{}l@{}}Wealth \& deprivation mapping \\Spending behavior profiling \\ Credit scoring\end{tabular} & \begin{tabular}[c]{@{}l@{}}\cite{Blumenstock2015,eagle2014,jean2016,Soto2011,venerandi2015}\\\cite{Singh2013} \\ \cite{sanpedro2015,Singh15}\end{tabular} \\ \hline
\textit{Public Health} & \begin{tabular}[c]{@{}l@{}}Epidemiologic studies\\ Environment and emergency mapping\\Mental Health\end{tabular} & \begin{tabular}[c]{@{}l@{}}\cite{FriasMartinez2011,tizzoni2014,Wesolowski2012}\\\cite{bengtsson2011,liu2013,lu2012} \\\cite{bogomolov2014stress,deChoudhury2013,deoliveira2011,faurholt14,lepri2016umuai,likamwa2013,matic16,osmani2015}\end{tabular} \\
\end{tabular}
\end{table}

\section{The dark side of data-driven decision-making for social good}
The potential positive impact of big data and machine learning-based approaches to decision-making is huge. However, several researchers and experts \cite{barocas2016,crawfordSchultz2014,pasquale2015,sweeney13,tufekci2015} have underlined what we refer to as \emph{the dark side} of data-driven decision-making, including violations of privacy, information asymmetry, lack of transparency, discrimination and social exclusion. In this section we turn our attention to these elements before outlining three key requirements that would be necessary in order to realize the positive impact, while minimizing the potential negative consequences of data-driven decision-making in the context of social good. 

\subsection{Computational violations of privacy}
Reports and studies \cite{federalcomm} have focused on the misuse of personal data disclosed by users and on the aggregation of data from different sources by entities playing as data brokers with direct implications in privacy. An often overlooked element is that the computational developments coupled with the availability of novel sources of behavioral data (\emph{e.g.} social media data, mobile phone data, etc.) now allow inferences about private information that may never have been disclosed. This element is essential to understand the issues raised by these algorithmic approaches.

A recent study by Kosinski \emph{et al.} \cite{kosinski2013private} combined data on Facebook ``Likes" and limited survey information to accurately predict a male user's sexual orientation, ethnic origin, religious and political preferences, as well as alcohol, drugs, and cigarettes use. Moreover, Twitter data has recently been used to identify people with a high likelihood of falling into depression before the onset of the clinical symptoms \cite{deChoudhury2013}.

It has also been shown that, despite the algorithmic advancements in anonymizing data, it is feasible to infer identities from anonymized human behavioral data, particularly when combined with information derived from additional sources. For example, Zang \emph{et al.} \cite{zang2011} have reported that if home and work addresses were available for some users, up to 35\% of users of the mobile network could be de-identified just using the two most visited towers, likely to be related to their home and work location. More recently, de Montjoye \emph{et al.} \cite{deMontjoye2013,deMontjoye2015} have demonstrated how unique mobility and shopping behaviors are for each individual. Specifically, they have shown that four spatio-temporal points are enough to uniquely identify 95\% of people in a mobile phone database of 1.5M people and to identify 90\% of people in a credit card database of 1M people.

\subsection{Information asymmetry and lack of transparency}
Both governments and companies use data-driven algorithms for decision making and optimization. Thus, accountability in government and corporate use of such decision making tools is fundamental in both validating their utility toward the public interest as well as redressing harms generated by these algorithms.

However, the ability to accumulate and manipulate behavioral data about customers and citizens on an unprecedented scale may give big companies and intrusive/authoritarian governments powerful means to manipulate segments of the population through targeted marketing efforts and social control strategies. In particular, we might witness an \emph{information asymmetry} situation where a powerful few have access and use knowledge that the majority do not have access to, thus leading to an --or exacerbating the existing-- asymmetry of power between the state or the big companies on one side and the people on the other side \cite{akerlof1970}. In addition, the nature and the use of various data-driven algorithms for social good, as well as the lack of computational or data literacy among citizens, makes algorithmic transparency difficult to generalize and accountability difficult to assess \cite{pasquale2015}.

Burrell \cite{burrell16} has provided a useful framework to characterize three different types of opacity in algorithmic decision-making: (1) \emph{intentional opacity}, whose objective is the protection of the intellectual property of the inventors of the algorithms. This type of opacity could be mitigated with legislation that would force decision-makers towards the use of open source systems. The new General Data Protection Regulations (GDPR) in the EU with a ``right to an explanation" starting in 2018 is an example of such legislation\footnote{Regulation (EU) 2016/679 of the European Parliament and of the Council of 27 April 2016 on the protection of natural persons with regard to the processing of personal data and on the free movement of such data, and repealing Directive 95/46/EC (General Data Protection Regulation) \url{http://eur-lex.europa.eu/eli/reg/2016/679/oj}}. However, there are clear corporate and governmental interests in favor of intentional opacity which make it difficult to eliminate this type of opacity; (2) \emph{illiterate opacity}, due to the fact that the vast majority of people lack the technical skills to understand the underpinnings of algorithms and machine learning models built from data. This kind of opacity might be attenuated with stronger education programs in computational thinking and by enabling that independent experts advice those affected by algorithm decision-making; and (3) \emph{intrinsic opacity}, which arises by the nature of certain machine learning methods that are difficult to interpret (\emph{e.g.} deep learning models). This opacity is well known in the machine learning community (usually referred to as the \emph{interpretability problem}). The main approach to combat this type of opacity requires using alternative machine learning models that are easy to interpret by humans, despite the fact that they might yield lower accuracy than black-box non-interpretable models. 

Fortunately, there is increasing awareness of the importance of reducing or eliminating the opacity of data-driven algorithmic decision-making systems. There are a number of research efforts and initiatives in this direction, including the Data Transparency Lab\footnote{http://www.datatransparencylab.org/} which is a ``community of technologists, researchers, policymakers and industry representatives working to advance online personal data transparency through research and design", and the DARPA Explainable Artificial Intelligence (XAI) project\footnote{http://www.darpa.mil/program/explainable-artificial-intelligence}. A tutorial on the subject has been held at the 2016 ACM Knowledge and Data Discovery conference~\cite{hajian2016algorithmic}. Researchers from New York University's Information Law Institute, such as Helen Nissenbaum and Solon Barocas, and Microsoft Research, such as Kate Crawford and Tarleton Gillespie, have held several workshops and conferences during the past few years on the ethical and legal challenges related to algorithmic governance and decision-making.\footnote{http://www.law.nyu.edu/centers/ili/algorithmsconference} A nominee for the National Book Award, Cathy O'Neil's book, ``Weapons of Math Destruction," details several case studies on harms and risks to public accountability associated with big data-driven algorithmic decision-making, particularly in the areas of criminal justice and education \cite{Oneil2016}. Recently, in partnership with Microsoft Research and others, the White House Office of Science and Technology Policy has co-hosted several public symposiums on the impacts and challenges of algorithms and artificial intelligence, specifically in social inequality, labor, healthcare and ethics.\footnote{https://www.whitehouse.gov/blog/2016/05/03/preparing-future-artificial-intelligence}

\subsection{Social exclusion and discrimination}
From a legal perspective, Tobler \cite{Tobler2008} argued that discrimination derives from ``the application of different rules or practices to comparable situations, or of the same rule or practice to different situations". In a recent paper, Barocas and Selbst \cite{barocas2016} elaborate that discrimination may be an artifact of the data collection and analysis process itself; more specifically, even with the best intentions, data-driven algorithmic decision-making can lead to discriminatory practices and outcomes. Algorithmic decision procedures can reproduce existing patterns of discrimination, inherit the prejudice of prior decision makers, or simply reflect the widespread biases that persist in society \cite{crawfordSchultz2014}. It can even have the perverse result of exacerbating existing inequalities by suggesting that historically disadvantaged groups actually deserve less favorable treatment \cite{Oneil2016}. 

Discrimination from algorithms can occur for several reasons. First, input data into algorithmic decisions may be poorly weighted, leading to \emph{disparate impact}; for example, as a form of \emph{indirect discrimination}, overemphasis of zip code within predictive policing algorithms can lead to the association of low-income African-American neighborhoods with areas of crime and as a result, the application of specific targeting based on group membership \cite{ChristinRosenblattBoyd2015}. Second, discrimination can occur from the decision to use an algorithm itself. Categorization -- through algorithmic classification, prioritization, association and filtering -- can be considered as a form of \emph{direct discrimination}, whereby algorithms are used for disparate treatment \cite{Diakopoulos2015}. Third, algorithms can lead to discrimination as a result of the misuse of certain models in different contexts \cite{calders2013}. Fourth, in a form of feedback loop, biased training data can be used both as evidence for the use of algorithms and as proof of their effectiveness \cite{calders2013}. 

The use of algorithmic data-driven decision processes may also result in individuals mistakenly being denied opportunities based not on their own action but on the actions of others with whom they share some characteristics. For example, some credit card companies have lowered a customer's credit limit, not based on the customer's payment history, but rather based on analysis of other customers with a poor repayment history that had shopped at the same establishments where the customer had shopped \cite{federalcomm}.

Indeed, we find increasing evidence of detrimental impact already taking place in current non-algorithmic approaches to credit scoring and generally to backgrounds checks. The latter have been widely used in recent years in several contexts: it is common to agree to be subjected to background checks when applying for a job, to lease a new apartment, etc. In fact, hundreds of thousands of people have unknowingly seen themselves adversely affected on existential matters such as job opportunities and housing availability due to simple but common mistakes (for instance, misidentification) in the procedures used by external companies to perform background checks\footnote{See, for instance, \url{http://www.chicagotribune.com/business/ct-background-check-penalties-1030-biz-20151029-story.html}}. It is worth noticing that the trivial procedural mistakes causing such adverse outcomes are bound to disappear once fully replaced with data-driven methodologies. Alas, this also means that should such methodologies not be transparent in their inner workings, the effects are likely to stay though with different roots. Further, the effort required to identify the causes of unfair and discriminative outcomes can be expected to be exponentially larger, as exponentially more complex will be the black-box models employed to assist in the decision-making process. This scenario highlights particularly well the need for machine learning models featuring transparency and accountability: adopting black-box approaches in scenarios where the lives of people would be seriously affected by a machine-driven decision could lead to forms of \emph{algorithmic stigma}\footnote{As a social phenomenon, the concept of stigma has received significant attention by sociologists, who under different frames highlighted and categorized the various factors leading individuals or groups to be discriminated by society, the countermoves often adopted by the stigmatized, and analyzed dynamics of reactions and evolution of stigma. We refer the interested reader to the review provided by Major and O'Brian \cite{major2005social}.}, a particularly creepy scenario considering that those stigmatized might never become aware of being so, and the stigmatizer will be an unaccountable machine. Recent advances in neural network-based (deep learning) models are yielding unprecedented accuracies in a variety of fields. However, such models tend to be difficult -- if not impossible -- to interpret, as previously explained. In this chapter, we highlight the need for data-driven machine learning models that are interpretable by humans when such models are going to be used to make decisions that affect individuals or groups of individuals.

\section{Requirements for positive disruption of data-driven policies}

As noted in the previous sections, both governments and companies are increasingly using data-driven algorithms for decision support and resource optimization. In the context of social good, accountability in the use of such powerful decision support tools is fundamental in both validating their utility toward the public interest as well as redressing corrupt or unjust harms generated by these algorithms. Several scholars have emphasized elements of what we refer to as the dark side of data-driven policies for social good, including violations of individual and group privacy, information asymmetry, lack of transparency, social exclusion and discrimination. Arguments against the use of social good algorithms typically call into question the use of machines in decision support and the need to protect the role of human decision-making. 

However, therein lies a huge potential and imperative for leveraging large scale human behavioral data to design and implement policies that would help improve the lives of millions of people. Recent debates have focused on characterizing data-driven policies as either ``good" or ``bad" for society. We focus instead on the potential of data-driven policies to lead to positive disruption, such that they reinforce and enable the powerful functions of algorithms as tools generating value while minimizing their dark side. 

In this  section, we present key \emph{human-centric} requirements for positive disruption, including a fundamental renegotiation of user-centric data ownership and management, the development of tools and participatory infrastructures towards increased algorithmic transparency and accountability, and the creation of living labs for experimenting and co-creating data-driven policies. We place humans at the center of our discussion as humans are ultimately both the actors and the subjects of the decisions made via algorithmic means. If we are able to ensure that these requirements are met, we should be able to realize the positive potential of data-driven algorithmic decision-making while minimizing the risks and possible negative unintended consequences.    

\subsection{User-centric data ownership and management}
A big question on the table for policy-makers, researchers, and intellectuals is: \emph{how do we unlock the value of human behavioral data while preserving the fundamental right to privacy?} This question implicitly recognizes the risks, in terms not only of possible abuses but also of a ``missed chance for innovation'', inherent to the current paradigm: the dominant \emph{siloed} approach to data collection, management, and exploitation, precludes participation to a wide range of actors, most notably to the very producers of personal data (\emph{i.e.} the users).

On this matter, new user-centric models for personal data management have been proposed, in
order to empower individuals with more control of their own data's life-cycle \cite{pentland2012}. To this end, researchers and companies are developing repositories which implement medium-grained
access control to different kinds of personally identifiable information (PII), such as
passwords, social security numbers and health data \cite{want2002}, location \cite{deMontjoye2014} and personal data collected by means of smartphones or connected devices \cite{deMontjoye2014}. A pillar of these approaches is represented by a \emph{Personal Data Eco-system}, composed by secure vaults of personal data whose owners are granted full control of.

Along this line, an interesting example is the Enigma platform \cite{zyskind2014} that leverages the recent technological trend of decentralization: advances in the fields of cryptography and decentralized computer networks have resulted in the emergence of a novel technology -- known as the \emph{blockchain} -- which has the potential to reduce the role of one of the most important actors in our society: the middle man \cite{benkler2006,defilippi2015}. By allowing people to transfer a unique piece of digital property or data to others, in a safe, secure, and immutable way, this technology can create digital currencies (\emph{e.g.} bitcoin) that are not backed by any governmental body \cite{nakamoto}; self-enforcing digital contracts, called smart contracts, whose execution does not require any human intervention (\emph{e.g.} Ethereum) \cite{szabo}; and decentralized marketplaces that aim to operate free from regulations \cite{defilippi2015}. Hence, Enigma tackles the challenge of providing a secure and trustworthy mechanism for the exchange of goods in a personal data market. To illustrate how the platform works, consider the following example: a group of data analysts of
an insurance company wishes to test a model that leverages people's mobile phone data.
Instead of sharing their raw data with the data analysts in the insurance company, the users can
securely store their data in Enigma, and only provide the data analysts with a permission to
execute their study. The data analysts are thus able to execute their code and obtain the
results, but nothing else. In the process, the users are compensated for having given access to
their data and the computers in the network are paid for their computing resources \cite{staiano2016}.

\subsection{Algorithmic transparency and accountability}
The deployment of a machine learning model entails a degree of \emph{trust} on how satisfactory its performance in the wild will be from the perspectives of both the builders and the users. Such trust is assessed at several points during an iterative model building process. Nonetheless, many of the state-of-the-art  machine learning-based models (\emph{e.g.} neural networks) act as black-boxes once deployed. When such models are used for decision-making, the lack of explanations regarding why and how they have reached their decisions poses several concerns. In order to address this limitation, recent research efforts in the machine learning community have proposed different approaches to make the algorithms more amenable to \emph{ex ante} and \emph{ex post} inspection. For example, a number of studies have attempted to tackle the issue of discrimination within algorithms by introducing tools to both identify \cite{berendt2014} and rectify \cite{calders2010,berendt2014,feldman2015} cases of unwanted bias. Recently, Ribeiro \emph{et al.} \cite{DBLP:conf/kdd/Ribeiro0G16} have proposed a model-agnostic method to derive explanations for the predictions of a given model.

An interesting ongoing initiative is the Open Algorithms (OPAL) project \footnote{\url{http://datapopalliance.org/open-algorithms-a-new-paradigm-for-using-private-data-for-social-good/}}, a multi-partner effort led by Orange, the MIT Media Lab, Data-Pop Alliance, Imperial College London, and the World Economic Forum, that aims to open –-without exposing-— data collected and stored by private companies by ``sending the code to the data" rather than the other way around. The goal is to enable the design, implementation and monitoring of development policies and programs, accountability of government action, and citizen engagement while leveraging the availability of large scale human behavioral data. OPAL's core will consist of an open platform allowing open algorithms to run on the servers of partner companies, behind their firewalls, to extract key development indicators and operational data of relevance for a wide range of potential users. 
Requests for approved, certified and pre-determined indicators by third parties --\emph{e.g.} mobility matrices, poverty maps, population densities-- will be sent to them via the platform; certified algorithms will run on the data in a multiple privacy-preserving manner, and results will be made available via an API. The platform will also be used to foster civic engagement of a broad range of social constituents --academic institutions, private sector companies, official institutions, non-governmental and civil society organizations. 
Overall, the OPAL initiative has three key objectives: (i) engage with data providers, users, and analysts at all the stages of algorithm development; (ii) contribute to building local capacities and help shaping the future technological, ethical and legal frameworks that will govern the collection, control and use of human behavioral data to foster social progress; and (iii) build data literacy among users and partners, conceptualized as ``the ability to constructively engage in society through and about data". Initiatives such as OPAL have the potential to enable more human-centric accountable and transparent data-driven decision-making and governance.  

\subsection{Living labs to experiment data-driven policies}
The use of real-time human behavioral data to design and implement policies has been traditionally outside the scope of the way of working in policy making. However, the potential of this type of data will only be realized when policy makers are able to analyze the data, to study human behavior and to test policies in the real world. A possible way is to build living laboratories –-communities of volunteers willing to try new ways of doing things in a natural setting-– in order to test ideas and hypotheses in a real life setting. An example is the Mobile Territorial Lab (MTL), a living lab launched by Fondazione Bruno Kessler, Telecom Italia, the MIT Media Lab and Telefonica, that has been observing the lives of more than 100 families through multiple channels for more than three years \cite{centellegher2016}. Data from multiple sources, including smartphones, questionnaires, experience sampling probes, etc. has been collected and used to create a multi-layered view of the lives of the study participants. In particular, social interactions (\emph{e.g.} call and SMS communications), mobility routines and spending patterns, etc. have been captured.
One of the MTL goals is to devise new ways of sharing personal data by means of Personal Data Store (PDS) technologies, in order to promote greater civic engagement. An example of an application enabled by PDS technologies is the sharing of best practices among families with young children. How do other families spend their money? How much do they get out and socialize? Once the individual gives permission, MyDataStore \cite{vescovi2014}, the PDS system used by MTL participants, allows such personal data to be collected, anonymized, and shared with other young families safely and automatically.

The MTL has been also used to investigate how to deal with the sensitivities of collecting and using deeply personal data in real-world situations. In particular, a MTL study investigated the perceived monetary value of mobile information and its association with behavioral characteristics and demographics; the results corroborate the arguments towards giving back to the people (users, citizens, according to the scenario) control on the data they constantly produce \cite{staiano2014money}.

Along these lines, Data-Pop Alliance and the MIT Media Lab launched in May 2016 a novel initiative called ``Laboratorio Urbano" in Bogot\'{a}, Colombia, in partnership with Bogot\'{a}'s city government and Chamber of Commerce. The main objective of the Bogot\'{a} Urban Laboratory is to contribute to the city's urban vitality, with a focus on mobility and safety, through collaborative research projects and dialogues involving the public and private sectors, academic institutions, and citizens. Similar initiatives are being planned in other major cities of the global south, including Dakar, Senegal, with the goal of strengthening and connecting local ecosystems where data-driven innovations can take place and scale. 

\begin{figure}    
    \centering
    \includegraphics[width=.6\linewidth]{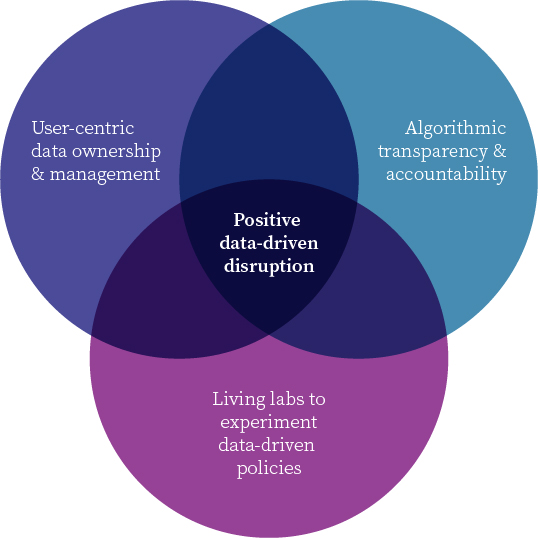}
    \caption{Requirements summary for positive data-driven disruption.}
    \label{fig:summary}
\end{figure}

Figure~\ref{fig:summary} provides the readers with a visual representation of the factors playing a significant role in positive data-driven disruption.

\section{Conclusion}
\label{sec:Y}
In this chapter we have provided an overview of both the opportunities and the risks of data-driven algorithmic decision-making for the public good. We are witnessing an unprecedented time in our history, where vast amounts of fine grained human behavioral data are available. The analysis of this data has the impact to help inform policies in public health, disaster management, safety, economic development and national statistics among others. In fact, the use of data is at the core of the 17 Sustainable Development Goals (SDGs) defined by United Nations, both in order to achieve the goals and to measure progress towards their achievement.  

While this is an exciting time for researchers and practitioners in this new field of computational social sciences, we need to be aware of the risks associated with these new approaches to decision making, including violation of privacy, lack of transparency, information asymmetry, social exclusion and discrimination. We have proposed three human-centric requirements that we consider to be of paramount importance in order to enable positive disruption of data-driven policy-making: user-centric data ownership and management; algorithmic transparency and accountability; and living labs to experiment with data-driven policies in the wild. It will be only when we honor these requirements that we will be able to move from the feared tyranny of data and algorithms to a data-enabled model of democratic governance running against tyrants and autocrats, and for the people. 




%
\bibliographystyle{plain}
\bibliography{TyrannyOfData}

\begin{thebibliography}{100}

\bibitem{akerlof1970}
G.A. Akerlof.
\newblock The market for ``lemons": Quality uncertainty and the market
  mechanism.
\newblock {\em The Quarterly Journal of Economics}, 84(3):488--500, 1970.

\bibitem{akerlof2009}
G.A. Akerlof and R.J. Shiller.
\newblock {\em Animal spirits: How human psychology drives the economy, and why
  it matters for global capitalism}.
\newblock Princeton University Press, 2009.

\bibitem{barocas2016}
S.~Barocas and A.D. Selbst.
\newblock Big data's disparate impact.
\newblock {\em California Law Review}, 104:671--732, 2016.

\bibitem{bengtsson2011}
L.~Bengtsson, X.~Lu, A.~Thorson, R.~Garfield, and J.~Von~Schreeb.
\newblock Improved response to disasters and outbreaks by tracking population
  movements with mobile phone network data: a post-earthquake geospatial study
  in haiti.
\newblock {\em PloS Medicine}, 8(8), 2011.

\bibitem{benkler2006}
Y.~Benkler.
\newblock {\em The wealth of networks}.
\newblock Yale University Press, New Haven, 2006.

\bibitem{berendt2014}
B.~Berendt and S.~Preibusch.
\newblock Better decision support through exploratory discrimination-aware data
  mining: Foundations and empirical evidence.
\newblock {\em Artificial Intelligence and Law}, 22(2):1572--8382, 2014.

\bibitem{Blondel15}
V.~D. Blondel, A.~Decuyper, and G.~Krings.
\newblock {A survey of results on mobile phone datasets analysis}.
\newblock {\em EPJ Data Science}, 4(10), 2015.

\bibitem{Blumenstock2015}
J.~Blumenstock, G.~Cadamuro, and R.~On.
\newblock Predicting poverty and wealth from mobile phone metadata.
\newblock {\em Science}, 350(6264):1073--1076, 2015.

\bibitem{bogomolov2014stress}
A.~Bogomolov, B.~Lepri, M.~Ferron, F.~Pianesi, and A.~Pentland.
\newblock Daily stress recognition from mobile phone data, weather conditions
  and individual traits.
\newblock In {\em Proceedings of the 22nd ACM International Conference on
  Multimedia}, pages 477--486. 2014.

\bibitem{bogomolov2015}
A.~Bogomolov, B.~Lepri, J.~Staiano, E.~Letouz{\'e}, N.~Oliver, F.~Pianesi, and
  A.~Pentland.
\newblock Moves on the street: Classifying crime hotspots using aggregated
  anonymized data on people dynamics.
\newblock {\em Big Data}, 3(3):148--158, 2015.

\bibitem{bogomolov2014}
A.~Bogomolov, B.~Lepri, J.~Staiano, N.~Oliver, F.~Pianesi, and A.~Pentland.
\newblock Once upon a crime: Towards crime prediction from demographics and
  mobile data.
\newblock In {\em Proceedings of the International Conference on Multimodal
  Interaction (ICMI)}, pages 427--434, 2014.

\bibitem{burrell16}
J.~Burrell.
\newblock How the machine {\textquoteleft}thinks{\textquoteright}:
  Understanding opacity in machine learning algorithms.
\newblock {\em Big Data \& Society}, 3(1), 2016.

\bibitem{calders2010}
T.~Calders and S.~Verwer.
\newblock Three naive bayes approaches for discrimination-free classification.
\newblock {\em Data Mining and Knowledge Discovery}, 21(2):277--292, 2010.

\bibitem{calders2013}
T.~Calders and I.~Zliobaite.
\newblock Why unbiased computational processes can lead to discriminative
  decision procedures.
\newblock In B.~Custers, T.~Calders, B.~Schermer, and T.~Zarsky, editors, {\em
  Discrimination and Privacy in the Information Society}, pages 43--57. 2013.

\bibitem{centellegher2016}
S.~Centellegher, M.~De~Nadai, M.~Caraviello, C.~Leonardi, M.~Vescovi,
  Y.~Ramadian, N.~Oliver, F.~Pianesi, A.~Pentland, F.~Antonelli, and B.~Lepri.
\newblock The mobile territorial lab: A multilayered and dynamic view on
  parents’ daily lives.
\newblock {\em EPJ Data Science}, 5(3), 2016.

\bibitem{Chainey2008}
S.P. Chainey, L.~Tompson, and S.~Uhlig.
\newblock The utility of hotspot mapping for predicting spatial patterns of
  crime.
\newblock {\em Security Journal}, 21:4--28, 2008.

\bibitem{ChristinRosenblattBoyd2015}
A.~Christin, A.~Rosenblatt, and d.~boyd.
\newblock Courts and predictive algorithms.
\newblock {\em Data \& Civil Rights Primer}, 2015.

\bibitem{citronPasquale2014}
D.K. Citron and F.~Pasquale.
\newblock The scored society.
\newblock {\em Washington Law Review}, 89(1):1--33, 2014.

\bibitem{crawfordSchultz2014}
K.~Crawford and J.~Schultz.
\newblock Big data and due process: Toward a framework to redress predictive
  privacy harms.
\newblock {\em Boston College Law Review}, 55(1):93--128, 2014.

\bibitem{deChoudhury2013}
M.~De~Choudhury, M.~Gamon, S.~Counts, , and E.~Horvitz.
\newblock Predicting depression via social media.
\newblock In {\em Proceedings of the 7th International AAAI Conference on
  Weblogs and Social Media}, 2013.

\bibitem{defilippi2015}
P.~De~Filippi.
\newblock The interplay between decentralization and privacy: The case of
  blockchain technologies.
\newblock {\em Journal of Peer Production}, 7, 2015.

\bibitem{deMontjoye2013}
Y.-A. de~Montjoye, C.~Hidalgo, M.~Verleysen, and V.~Blondel.
\newblock Unique in the crowd: The privacy bounds of human mobility.
\newblock {\em Scientific Reports}, 3, 2013.

\bibitem{deMontjoye2015}
Y.-A. de~Montjoye, L.~Radaelli, V.K. Singh, and A.~Pentland.
\newblock Unique in the shopping mall: On the re-identifiability of credit card
  metadata.
\newblock {\em Science}, 347(6221):536--539, 2015.

\bibitem{deMontjoye2014}
Y.-A. de~Montjoye, E.~Shmueli, S.~Wang, and A.~Pentland.
\newblock Openpds: Protecting the privacy of metadata through safeanswers.
\newblock {\em PloS One}, (10.1371), 2014.

\bibitem{deoliveira2011}
R.~de~Oliveira, A.~Karatzoglou, P.~Concejero~Cerezo, A.~Armenta Lopez~de
  Vicu{\~n}a, and N.~Oliver.
\newblock Towards a psychographic user model from mobile phone usage.
\newblock In {\em CHI'11 Extended Abstracts on Human Factors in Computing
  Systems}, pages 2191--2196. ACM, 2011.

\bibitem{devarajan2013}
S.~Devarajan.
\newblock Africa's statistical tragedy.
\newblock {\em Review of Income and Wealth}, 59(S1):S9--S15, 2013.

\bibitem{Diakopoulos2015}
N.~Diakopoulos.
\newblock Algorithmic accountability: Journalistic investigation of
  computational power structures.
\newblock {\em Digital Journalism}, 2015.

\bibitem{Easterly2014}
W.~Easterly.
\newblock {\em {The Tyranny of Experts}}.
\newblock Basic Books, 2014.

\bibitem{Eck2005}
J.~Eck, S.~Chainey, J.~Cameron, and R.~Wilson.
\newblock Mapping crime: understanding hotspots.
\newblock {\em National Institute of Justice: Washington DC}, 2005.

\bibitem{faurholt14}
M.~Faurholt-Jepsena, M.~Frostb, M.~Vinberga, E.M. Christensena, J.E. Bardram,
  and L.V. Kessinga.
\newblock Smartphone data as objective measures of bipolar disorder symptoms.
\newblock {\em Psychiatry Research}, 217:124--127, 2014.

\bibitem{feldman2015}
M.~Feldman, S.A. Friedler, J.~Moeller, C.~Scheidegger, and
  S.~Venkatasubramanian.
\newblock Certifying and removing disparate impact.
\newblock In {\em Proceedings of the 21th ACM SIGKDD International Conference
  on Knowledge Discovery and Data Mining}, pages 259--268, 2015.

\bibitem{ferguson2012}
A.G. Ferguson.
\newblock Crime mapping and the fourth amendment: Redrawing high-crime areas.
\newblock {\em Hastings Law Journal}, 63:179--232, 2012.

\bibitem{fields1989}
G.~Fields.
\newblock Changes in poverty and inequality.
\newblock {\em World Bank Research Observer}, 4:167--186, 1989.

\bibitem{fiske1998}
S.T. Fiske.
\newblock Stereotyping, prejudice, and discrimination.
\newblock In D.T. Gilbert, S.T. Fiske, and G.~Lindzey, editors, {\em Handbook
  of Social Psychology}, pages 357--411. Boston: McGraw-Hill, 1998.

\bibitem{FriasMartinez2011}
E.~Frias-Martinez, G.~Williamson, and V.~Frias-Martinez.
\newblock An agent-based model of epidemic spread using human mobility and
  social network information.
\newblock In {\em Social Computing (SocialCom), 2011 International Conference
  on}, pages 57--64. IEEE, 2011.

\bibitem{Gillespie2014}
T.~Gillespie.
\newblock The relevance of algorithms.
\newblock In T.~Gillespie, P.~Boczkowski, and K.~Foot, editors, {\em Media
  technologies: Essays on communication, materiality, and society}, pages
  167--193. MIT Press, 2014.

\bibitem{ginsberg2009}
J.~Ginsberg, M.H. Mohebbi, R.S. Patel, L.~rammer, M.S. Smolinski, and
  L.~Brilliant.
\newblock Detecting influenza epidemics using search engine query data.
\newblock {\em Nature}, 457:1012--1014, 2009.

\bibitem{hajian2016algorithmic}
Sara Hajian, Francesco Bonchi, and Carlos Castillo.
\newblock Algorithmic bias: From discrimination discovery to fairness-aware
  data mining.
\newblock In {\em Proceedings of the 22nd ACM SIGKDD International Conference
  on Knowledge Discovery and Data Mining}, pages 2125--2126. ACM, 2016.

\bibitem{jean2016}
N.~Jean, M.~Burke, M.~Xie, W.M. Davis, D.B. Lobell, and S.~Ermon.
\newblock Combining satellite imagery and machine learning to predict poverty.
\newblock {\em Science}, 353(6301):790--794, 2016.

\bibitem{jerven2013}
M.~Jerven.
\newblock {\em Poor numbers: How we are misled by african development
  statistics and what to do about it}.
\newblock Cornell University Press, 2013.

\bibitem{king2011ensuring}
G.~King.
\newblock Ensuring the data-rich future of the social sciences.
\newblock {\em Science}, 2011.

\bibitem{kosinski2013private}
M.~Kosinski, D.~Stillwell, and T.~Graepel.
\newblock Private traits and attributes are predictable from digital records of
  human behavior.
\newblock {\em Proceedings of the National Academy of Sciences},
  110(15):5802--5805, 2013.

\bibitem{kuznets1955}
S.~Kuznets.
\newblock Economic growth and income inequality.
\newblock {\em American Economic Review}, 45:1--28, 1955.

\bibitem{Latzer2015}
M.~Latzer, K.~Hollnbuchner, N.~Just, and F.~Saurwein.
\newblock The economics of algorithmic selection on the internet.
\newblock In J.~Bauer and M.~Latzer, editors, {\em Handbook on the Economics of
  the Internet}. Edward Elgar, Cheltenham, Northampton, 2015.

\bibitem{Lazer09}
D.~Lazer, A.~Pentland, L.~Adamic, S.~Aral, A-L. Barabasi, D.~Brewer,
  N.~Christakis, N.~Contractor, J.~Fowler, M.~Gutmann, T.~Jebara, G.~King,
  M.~Macy, D.~Roy, and M.~Van~Alstyne.
\newblock Computational social science.
\newblock {\em Science}, 323(5915):721--723, 2009.

\bibitem{lepri2016umuai}
B.~Lepri, J.~Staiano, E.~Shmueli, F.~Pianesi, and A.~Pentland.
\newblock The role of personality in shaping social networks and mediating
  behavioral change.
\newblock {\em User Modeling and User-Adapted Interaction}, 26(2):143--175,
  2016.

\bibitem{likamwa2013}
R.~LiKamWa, Y.~Liu, N.D. Lane, and L.~Zhong.
\newblock Moodscope: Building a mood sensor from smartphone usage patterns.
\newblock In {\em Proceedings of the 11th Annual International Conference on
  Mobile Systems, Applications, and Service (MobiSys)}, pages 389--402. 2013.

\bibitem{liu2013}
H.Y. Liu, E.~Skjetne, and M.~Kobernus.
\newblock Mobile phone tracking: In support of modelling traffic-related air
  pollution contribution to individual exposure and its implications for public
  health impact assessment.
\newblock {\em Environmental Health}, 12, 2013.

\bibitem{eagle2014}
T.~Louail, M.~Lenormand, O.~G.~Cantu Ros, M.~Picornell, R.~Herranz,
  E.~Frias-Martinez, J.~J. Ramasco, and M.~Barthelemy.
\newblock {From mobile phone data to the spatial structure of cities}.
\newblock {\em Scientific Reports}, 4(5276), Jun 2014.

\bibitem{lu2012}
X.~Lu, L.~Bengtsson, and P.~Holme.
\newblock Predictability of population displacement after the 2010 haiti
  earthquake.
\newblock {\em Proceedings of the National Academy of Sciences}, 109:11576--81,
  2012.

\bibitem{major2005social}
B.~Major and L.T. O'Brien.
\newblock The social psychology of stigma.
\newblock {\em Annual Review of Psychology}, 56:393--421, 2005.

\bibitem{matic16}
A.~Matic and N.~Oliver.
\newblock The untapped opportunity of mobile network data for mental health.
\newblock In {\em Future of Pervasive Health Workshop}. ACM, 6 2016.

\bibitem{Mohler2011}
G.O. Mohler, M.B. Short, P.J. Brantingham, F.P. Schoenberg, and G.E. Tita.
\newblock Self-exciting point process modeling of crime.
\newblock {\em Journal of the American Statistical Association},
  (106):100--108, 2011.

\bibitem{nakamoto}
S.~Nakamoto.
\newblock Bitcoin: A peer-to-peer electronic cash system.
\newblock Technical report, Kent University, 2009.

\bibitem{ofli2016}
F.~Ofli, P.~Meier, M.~Imran, C.~Castillo, D.~Tuia, N.~Rey, J.~Briant,
  P.~Millet, F.~Reinhard, M.~Parkan, and S.~Joost.
\newblock Combining human computing and machine learning to make sense of big
  (aerial) data for disaster response.
\newblock {\em Big Data}, 4:47--59, 2016.

\bibitem{ohm2010}
P.~Ohm.
\newblock Broken promises of privacy: Responding to the surprising failure of
  anonymization.
\newblock {\em UCLA Law Review}, 57:1701--1777, 2010.

\bibitem{oliver15}
N.~Oliver, A.~Matic, and E.~Frias-Martinez.
\newblock Mobile network data for public health: Opportunities and challenges.
\newblock {\em Frontiers in Public Health}, 3:189, 2015.

\bibitem{Oneil2016}
C.~O'Neil.
\newblock {\em Weapons of math destruction: How big data increases inequality
  and threatens democracy}.
\newblock Crown, 2016.

\bibitem{osmani2015}
V.~Osmani, A.~Gruenerbl, G.~Bahle, Lukowicz~P. Haring, C., and Mayora O.
\newblock Smartphones in mental health: Detecting depressive and manic
  episodes.
\newblock {\em IEEE Pervasive Computing}, 14(3):10--13, 2015.

\bibitem{pager2008}
D.~Pager and H.~Shepherd.
\newblock The sociology of discrimination: Racial discrimination in employment,
  housing, credit and consumer market.
\newblock {\em Annual Review of Sociology}, 34:181--209, 2008.

\bibitem{pasquale2015}
F.~Pasquale.
\newblock {\em The Black Blox Society: The secret algorithms that control money
  and information}.
\newblock Harvard University Press, 2015.

\bibitem{pastor2014}
D.~Pastor-Escuredo, Y.~Torres~Fernandez, J.M. Bauer, A.~Wadhwa,
  C.~Castro-Correa, L.~Romanoff, J.G. Lee, A.~Rutherford, V.~Frias-Martinez,
  N.~Oliver, Frias-Martinez E., and M.~Luengo-Oroz.
\newblock Flooding through the lens of mobile phone activity.
\newblock In {\em IEEE Global Humanitarian Technology Conference, GHTC'14}.
  IEEE, 2014.

\bibitem{pentland2012}
A.~Pentland.
\newblock Society's nervous system: Building effective government, energy, and
  public health systems.
\newblock {\em IEEE Computer}, 45(1):31--38, 2012.

\bibitem{perry2013}
W.L. Perry, B.~McInnis, C.C. Price, S.C. Smith, and J.S. Hollywood.
\newblock {\em Predictive policing: The role of crime forecasting in law
  enforcment operations}.
\newblock Rand Corporation, 2013.

\bibitem{whitehouse}
J.~Podesta, P.~Pritzker, E.J. Moniz, J.~Holdren, and J.~Zients.
\newblock Big data: Seizing opportunities, preserving values.
\newblock Technical report, Executive Office of the President, 2014.

\bibitem{federalcomm}
E.~Ramirez, J.~Brill, M.K. Ohlhausen, and T.~McSweeny.
\newblock Big data: A tool for inclusion or exclusion?
\newblock Technical report, Federal Trade Commission, January 2016.

\bibitem{ratcliffe2006}
J.H. Ratcliffe.
\newblock A temporal constraint theory to explain opportunity-based spatial
  offending patterns.
\newblock {\em Journal of Research in Crime and Delinquency}, 43(3):261--291,
  2006.

\bibitem{ravallion2016}
M.~Ravallion.
\newblock {\em The economics of poverty: History, measurement, and policy}.
\newblock Oxford University Press, 2016.

\bibitem{DBLP:conf/kdd/Ribeiro0G16}
M.T. Ribeiro, S.~Singh, and C.~Guestrin.
\newblock "why should {I} trust you?": Explaining the predictions of any
  classifier.
\newblock In {\em Proceedings of the 22nd {ACM} {SIGKDD} International
  Conference on Knowledge Discovery and Data Mining, San Francisco, CA, USA,
  August 13-17, 2016}, pages 1135--1144, 2016.

\bibitem{samuelson1988}
W.~Samuelson and R.~Zeckhauser.
\newblock Status quo bias in decision making.
\newblock {\em Journal of Risk and Uncertainty}, (1):7--59, 1988.

\bibitem{sanpedro2015}
J.~San~Pedro, D.~Proserpio, and N.~Oliver.
\newblock Mobiscore: Towards universal credit scoring from mobile phone data.
\newblock In {\em Proceedings of the International Conference on User Modeling,
  Adaptation and Personalization (UMAP)}, pages 195--207, 2015.

\bibitem{Short2008}
M.~B. Short, M.~R. D'Orsogna, V.~B. Pasour, G.~E. Tita, P.~J. Brantingham,
  A.~L. Bertozzi, and L.~B. Chayes.
\newblock A statistical model of criminal behavior.
\newblock {\em Mathematical Models and Methods in Applied Sciences},
  18(supp01):1249--1267, 2008.

\bibitem{Singh15}
V.~K. Singh, B.~Bozkaya, and A.~Pentland.
\newblock {Money walks: Implicit mobility behavior and financial well-being}.
\newblock {\em PLOS ONE}, 10(8):e0136628, 2015.

\bibitem{Singh2013}
V.K. Singh, L.~Freeman, B.~Lepri, and A.~Pentland.
\newblock Predicting spending behavior using socio-mobile features.
\newblock In {\em Social Computing (SocialCom), 2013 International Conference
  on}, pages 174--179. IEEE, 2013.

\bibitem{Capra2014}
C.~Smith-Clarke, A.~Mashhadi, and L.~Capra.
\newblock Poverty on the cheap: Estimating poverty maps using aggregated mobile
  communication networks.
\newblock In {\em Proceedings of the 32nd ACM Conference on Human Factors in
  Computing Systems (CHI2014)}, 2014.

\bibitem{Soto2011}
V.~Soto, V.~Frias-Martinez, J.~Virseda, and E.~Frias-Martinez.
\newblock Prediction of socioeconomic levels using cell phone records.
\newblock In {\em Proceedings of the International conference on UMAP}, pages
  377--388, 2011.

\bibitem{staiano2014money}
J.~Staiano, N.~Oliver, B.~Lepri, R.~de~Oliveira, M.~Caraviello, and N.~Sebe.
\newblock Money walks: a human-centric study on the economics of personal
  mobile data.
\newblock In {\em Proceedings of the 2014 ACM International Joint Conference on
  Pervasive and Ubiquitous Computing}, pages 583--594. ACM, 2014.

\bibitem{staiano2016}
J.~Staiano, G.~Zyskind, B.~Lepri, N.~Oliver, and A.~Pentland.
\newblock The rise of decentralized personal data markets.
\newblock In D.~Shrier and A.~Pentland, editors, {\em Trust::Data: A New
  Framework for Identity and Data Sharing}. CreateSpace Independent Publishing
  Platform, 2016.

\bibitem{sweeney13}
L.~Sweeney.
\newblock Discrimination in online ad delivery.
\newblock {\em Available at SSRN: http://ssrn.com/abstract=2208240}, 2013.

\bibitem{szabo}
N.~Szabo.
\newblock Formalizing and securing relationships on public networks.
\newblock {\em First Monday}, 2(9), 1997.

\bibitem{thomas2009}
L.~Thomas.
\newblock {\em Consumer credit models: Pricing, profit, and portfolios}.
\newblock New York: Oxford University Press, 2009.

\bibitem{tizzoni2014}
M.~Tizzoni, P.~Bajardi, A.~Decuyper, G.~Kon Kam~King, C.M. Schneider,
  V.~Blondel, Z.~Smoreda, M.C. Gonzalez, and V.~Colizza.
\newblock On the use of human mobility proxies for modeling epidemics.
\newblock {\em PLoS Computational Biology}, 10(7), 2014.

\bibitem{Tobler2008}
C.~Tobler.
\newblock Limits and potential of the concept of indirect discrimination.
\newblock Technical report, European Network of Legal Experts in
  Anti-Discrimination, 2008.

\bibitem{Toole2011}
J.L. Toole, N.~Eagle, and J.B. Plotkin.
\newblock Spatiotemporal correlations in criminal offense records.
\newblock {\em ACM Transactions on Intelligent Systems and Technology},
  2(4):38:1--38:18, July 2011.

\bibitem{traunmueller2014}
M.~Traunmueller, G.~Quattrone, and L.~Capra.
\newblock Mining mobile phone data to investigate urban crime theories at
  scale.
\newblock In {\em Proceedings of the International Conference on Social
  Informatics}, pages 396--411, 2014.

\bibitem{tufekci2015}
Z.~Tufekci.
\newblock Algorithmic harms beyond facebook and google: Emergent challenges of
  computational agency.
\newblock {\em Colorado Technology Law Journal}, 13:203--218, 2015.

\bibitem{tversky1974}
A.~Tverksy and D.~Kahnemann.
\newblock Judgment under uncertainty: Heuristics and biases.
\newblock {\em Science}, 185(4157):1124--1131, 1974.

\bibitem{venerandi2015}
A.~Venerandi, G.~Quattrone, L.~Capra, D.~Quercia, and D.~Saez-Trumper.
\newblock Measuring urban deprivation from user generated content.
\newblock In {\em Proceedings of the 18th ACM Conference on Computer Supported
  Cooperative Work \& Social Computing (CSCW2015)}, 2015.

\bibitem{vescovi2014}
M.~Vescovi, C.~Perentis, C.~Leonardi, B.~Lepri, and C.~Moiso.
\newblock My data store: Toward user awareness and control on personal data.
\newblock In {\em Proceedings of the 2014 ACM International Joint Conference on
  Pervasive and Ubiquitous Computing: Adjunct Publication}, pages 179--182,
  2014.

\bibitem{wang2016}
H.~Wang, Z.~Li, D.~Kifer, and C.~Graif.
\newblock Crime rate inference with big data.
\newblock In {\em Proceedings of International conference on KDD}, 2016.

\bibitem{wang2013}
T.~Wang, C.~Rudin, D.~Wagner, and R.~Sevieri.
\newblock Learning to detect patterns of crime.
\newblock In {\em Machine Learning and Knowledge Discovery in Databases}, pages
  515--530. Springer, 2013.

\bibitem{want2002}
R.~Want, T.~Pering, G.~Danneels, M.~Kumar, M.~Sundar, and J.~Light.
\newblock The personal server: Changing the way we think about ubiquitous
  computing.
\newblock In {\em Proceedings of 4th International Conference on Ubiquitous
  Computing}, pages 194--209, 2002.

\bibitem{Weisburd2008}
D.~Weisburd.
\newblock Place-based policing.
\newblock {\em Ideas in American Policing}, 9:1--16, 2008.

\bibitem{Wesolowski2012}
A.~Wesolowski, N.~Eagle, A.~Tatem, D.~Smith, R.~Noor, and C.~Buckee.
\newblock Quantifying the impact of human mobility on malaria.
\newblock {\em Science}, 338(6104):267--270, 2012.

\bibitem{Wesolowski2014}
A.~Wesolowski, G.~Stresman, N.~Eagle, J.~Stevenson, C.~Owaga, E.~Marube,
  T.~Bousema, C.~Drakeley, J.~Cox, and C.O. Buckee.
\newblock Quantifying travel behavior for infectious disease research: A
  comparison of data from surveys and mobile phones.
\newblock {\em Scientific Reports}, 4, 2014.

\bibitem{Willson2016}
M.~Willson.
\newblock Algorithms (and the) everyday.
\newblock {\em Information, Communication \& Society}, 2016.

\bibitem{Wilson16}
R.~Wilson, E.~Erbach-Schoenengerg, M.~Albert, D.~Power, Tudge S., and
  Gonzalez~M. et~al.
\newblock {Rapid and Near Real-time Assessments of Population Displacement
  Using Mobile Phone Data Following Disasters: The 2015 Nepal Earthquake}.
\newblock {\em PLOS Current Disasters}, February 2016.

\bibitem{zang2011}
H.~Zang and J.~Bolot.
\newblock Anonymization of location data does not work: A large-scale
  measurement study.
\newblock In {\em Proceedings of 17th ACM Annual International Conference on
  Mobile Computing and Networking}, pages 145--156, 2011.

\bibitem{zarsky2016}
T.~Zarsky.
\newblock The trouble with algorithmic decisions: An analytic road map to
  examine efficiency and fairness in automated and opaque decision making.
\newblock {\em Science, Technology, and Human Values}, 41(1):118--132, 2016.

\bibitem{zarsky2012}
T.Z. Zarsky.
\newblock Automated prediction: Perception, law and policy.
\newblock {\em Communications of the ACM}, 4:167--186, 1989.

\bibitem{zyskind2014}
G.~Zyskind, O.~Nathan, and A.~Pentland.
\newblock Decentralizing privacy: Using blockchain to protect personal data.
\newblock In {\em Proceedings of IEEE Symposium on Security and Privacy
  Workshops}, pages 180--184. 2014.

\end{thebibliography}

\end{document}